\documentclass[aps,pra,preprint,groupedaddress]{revtex4-1}
\usepackage{amsmath}
\usepackage{graphicx}

\begin{document}


\title{New mechanism for polarization singularity of vector vortex beams}


\author{Chun-Fang Li}
\email[]{cfli@shu.edu.cn}
\affiliation{Department of Physics, Shanghai University, 99 Shangda Road, 200444 Shanghai, China}


\date{\today}

\begin{abstract}
It was recently realized that the polarization bases of the plane-wave modes in the integral representation of a light beam need to be determined by a degree of freedom arising from the divergence-free Maxwell's equation. This is a frequently introduced real unit vector in the literature, called Stratton vector. The polarization bases so determined are singular at the momentum that is parallel to the Stratton vector.
Here we show that the polarization singularity of vector vortex beams given by the integral representation comes from the singularity of the polarization bases in association with the Stratton vector.
The consistency of the polarization structure of the vector vortex beams with their polarization bases is also discussed.
\end{abstract}


\maketitle


\newpage

\section{Introduction}

Vector vortex (VV) beams have drawn more and more attention. They refer to light beams the polarization of which varies over the transverse profile \cite{Hall}. Of these, cylindrical-vector beams \cite{Pohl, Jord-H, Youn-B, Zhan} have a rotationally invariant polarization. A key property of VV beams is their polarization singularity (PS) \cite{Bomz-BKH}, a very common phenomenon of light \cite{Denn-OP}. Usually, a VV beam is viewed as a superposition of two polarization-orthogonal scalar beams \cite{Bomz-BKH, McLa-KF, D'Am-CG, Khaj-G}. Because scalar beams do not have polarization singularities \cite{Denn-OP}, it seems that the superposition of two polarization-orthogonal scalar beams is the only mechanism for the PS of VV beams. The purpose of this Rapid Communication is to put forward a completely different mechanism for the PS of VV beams.

As is well known, the electric field of a light beam can always be represented by an integral with respect to the wavevector. It was discovered not long ago \cite{Li08, Li09} that the polarization bases associated with the wavevector, which are not unique \cite{Mand-W}, can be determined by a degree of freedom, a frequently introduced real unit vector in the literature \cite{Stra, Gree-W, Patt-A, Davi-P, Enk-N}.
So determined polarization bases are singular \cite{Yang-L} at the wavevector that is parallel to the degree of freedom.
We will show that the integral representation gives VV beams when the degree of freedom is parallel to the propagation axis, on one hand, and such a VV beam is not at all a superposition of two polarization-orthogonal scalar beams, on the other hand.
By this it is meant that the singularity of the polarization bases provides a new mechanism for the PS of VV beams.

Let us begin with the integral representation \cite{Li08, Li09}. The complex-valued electric field of a free-space light beam propagating along the $z$ axis is represented by the integral with respect to the wavevector,
\begin{equation}\label{FT}
    \mathbf{E} (\mathbf{x},t)
   =\frac{1}{(2 \pi)^{3/2}} \int \mathbf{a}(\mathbf{k}) e (\mathbf{k})
     \exp[i(\mathbf{k} \cdot \mathbf{x}-\omega t)] d^3 k ,
\end{equation}
where $\omega=c k$, $c$ is the speed of light in free space, $k=|\mathbf{k}|$, the scalar function $e(\mathbf{k})$ stands for the strength of the electric field in momentum space, and the unit-vector function $\mathbf{a} (\mathbf{k})$ obeying
$|\mathbf{a} (\mathbf{k})|=1$
is the polarization vector \cite{Akhi-B, Cohe-DG}.
The polarization vector at each wavevector is expanded in terms of a pair of orthonormal polarization bases $\mathbf u$ and $\mathbf v$,
\begin{equation}\label{QUT}
\mathbf{a}(\mathbf{k})=\alpha_u \mathbf{u} +\alpha_v \mathbf{v} \equiv \varpi \alpha,
\end{equation}
where
$
\varpi =(
         \begin{array}{cc}
            \mathbf{u} & \mathbf{v} \\
         \end{array}
        )
$
is a 3-by-2 matrix that contains the polarization bases as its column vectors and
$
\alpha =\bigg(\begin{array}{c}
                \alpha_u \\
                \alpha_v
              \end{array}
        \bigg)
$
is the Jones vector obeying
$\alpha^\dag \alpha=1$.
The polarization bases associated with the wavevector are determined by a real unit vector $\mathbf I$ \cite{Stra, Gree-W, Patt-A, Davi-P, Enk-N} in the following way,
\begin{equation}\label{PB}
    \mathbf{u} =\mathbf{v} \times \frac{\mathbf k}{k},                             \quad
    \mathbf{v} =\frac{\mathbf{I} \times \mathbf{k}}{|\mathbf{I} \times \mathbf{k}|}.
\end{equation}
Since such a vector was first introduced by Stratton \cite{Stra}, we will refer to it as the Stratton vector.
We stress that in contrast with the solution of vector paraxial wave equation \cite{Hall}, the electric field given by Eqs. (\ref{FT})-(\ref{PB}) is an exact solution of free-space Maxwell's equations regardless of the Stratton vector, the Jones vector, and the strength factor. It satisfies not only the accurate vector wave equation
\begin{equation}\label{WE}
    \nabla^2 \mathbf{E}-\frac{1}{c^2} \frac{\partial^2 \mathbf{E}}{\partial t^2} =0,
\end{equation}
but also the Maxwell's equation,
\begin{equation}\label{ME}
    \nabla \cdot \mathbf{E}=0.
\end{equation}
In fact, it is from the divergence-free equation (\ref{ME}) that the Stratton vector stems \cite{Li08, Li09}.
Obviously, the polarization bases (\ref{PB}) are indeterminate or singular at the wavevector that is parallel to the Stratton vector.
Let us first show that if the Stratton vector is parallel to the propagation axis, proper Jones vector and strength factor will give rise to VV beams.

\section{Vector vortex beams and origin of their polarization singularity}

To this end, we consider the Stratton vector along the propagation axis,
$\mathbf{I}^\parallel = \bar z$, where $\bar z$ is the unit vector along the $z$ axis. The wavevector-associated polarization bases it determines take the form of
\begin{equation}\label{PB-para}
  \mathbf{u}^\parallel =\bar{\vartheta} =\bar{\rho} \cos \vartheta -\bar{z} \sin \vartheta , \quad
  \mathbf{v}^\parallel =\bar{\varphi},
\end{equation}
where $\vartheta$ is the polar angle of the wavevector in spherical coordinates, $\bar \vartheta$ denotes the unit vector along the polar direction, and $\bar \rho$ and $\bar \varphi$ are, respectively, the unit vectors along the radial and azimuthal directions in cylindrical coordinates.
They are singular on the propagation axis, $\sin \vartheta =0$.
In addition, we choose the Jones vector to be
$
\alpha_{1+}=\bigg(\begin{array}{c}
                    1 \\
                    0
                  \end{array}
            \bigg)
$
and
$
\alpha_{1-}=\bigg(\begin{array}{c}
                    0 \\
                    1
                  \end{array}
            \bigg)
$,
the eigenfunctions of the Pauli matrix
$
\hat{\sigma}_1 =\bigg(\begin{array}{cc}
                        1 &  0 \\
                        0 & -1
                      \end{array}
                \bigg)
$
with eigenvalues $\sigma_{1+} =1$ and $\sigma_{1-} =-1$, respectively. The polarization vectors described by $\alpha_{1+}$ and $\alpha_{1-}$ are
\begin{subequations}\label{PV-VVB}
\begin{align}
  \mathbf{a}_{1+}^\parallel =\varpi^\parallel \alpha_{1+} &
 =\bar{\rho} \cos \vartheta -\bar{z} \sin \vartheta ,        \label{PV-TM} \\
  \mathbf{a}_{1-}^\parallel =\varpi^\parallel \alpha_{1-} &
 =\bar{\varphi},                                             \label{PV-TE}
\end{align}
\end{subequations}
respectively, where
$
\varpi^\parallel =(\begin{array}{cc}
                     \mathbf{u}^\parallel & \mathbf{v}^\parallel
                   \end{array}
                  )
$.
Finally, without loss of generality, we consider monochromatic light beams of the following strength factor,
\begin{equation}\label{OAMB}
    e_m (\mathbf{k})= \frac{\sqrt{2 \pi}}{i k^2} \delta(k-k_0 ) f(\vartheta) \exp(i m \varphi),
\end{equation}
where $f(\vartheta)$ is any physically allowed function, $\varphi$ is the azimuthal angle of the wavevector, $m$ is an integer, and the additional factor $\frac{\sqrt{2 \pi}}{i k^2}$ is introduced for later convenience.
Upon inserting Eqs. (\ref{PV-VVB}) and (\ref{OAMB}) into Eq. (\ref{FT}) and making use of the identity,
\begin{equation}\label{iden}
    \exp(i \xi \cos \psi)= \sum_{n=- \infty}^{\infty} i^n J_n (\xi) \exp(i n \psi),
\end{equation}
we get
\begin{subequations}\label{VVB}
\begin{align}
  \mathbf{E}^\parallel_{1+,m} &
 =\frac{i^m}{2} e^{i m \phi}
   [(A_{m+1}-A_{m-1}) \bar{r} -i (A_{m+1}+A_{m-1}) \bar{\phi} +2i B_m \bar{z}], \label{VVB1} \\
  \mathbf{E}^\parallel_{1-,m} &
 =\frac{i^m}{2} e^{i m \phi}
   [(A'_{m+1}-A'_{m-1}) \bar{\phi} +i (A'_{m+1}+A'_{m-1}) \bar{r}],             \label{VVB2}
\end{align}
\end{subequations}
where a time-dependent factor $\exp(-i \omega_0 t)$ is assumed, $\omega_0 =c k_0$,
\begin{eqnarray}
  A_m (r,z) &=& \int_0^\pi  f(\vartheta) J_m (k_\perp r) \exp(i k_\parallel z)
                            \cos \vartheta \sin \vartheta d \vartheta ,        \nonumber \\
  B_m (r,z) &=& \int_0^\pi  f(\vartheta) J_m (k_\perp r) \exp(i k_\parallel z)
                            \sin^2 \vartheta d \vartheta ,                     \nonumber \\
  A'_m (r,z) &=& \int_0^\pi f(\vartheta) J_m (k_\perp r) \exp(i k_\parallel z)
                            \sin \vartheta d \vartheta ,                       \label{A'm}
\end{eqnarray}
$k_\perp =k_0 \sin \vartheta$, $k_\parallel =k_0 \cos \vartheta$, $(r, \phi, z)$ are cylindrical coordinates of the position vector $\mathbf x$, $\bar r$ and $\bar \phi$ are, respectively, the radial and azimuthal unit vectors in position space, and $J_m $ is the Bessel function of the first kind. Apparently, they are mutually orthogonal VV beams.
According to the principle of duality \cite{Davi-P81}, VV beam (\ref{VVB1}) is a TM mode and VV beam (\ref{VVB2}) a TE mode.
When $m=0$, they reduce to the following cylindrical-vector beams,
\begin{equation*}
    \mathbf{E}^\parallel_{1+,0} =A_1 \bar{r} +i B_0 \bar{z}, \quad
    \mathbf{E}^\parallel_{1-,0} =A'_1 \bar{\phi}.
\end{equation*}

Corresponding to the singularity of polarization bases (\ref{PB-para}) on the propagation axis, the unit vectors $\bar r$ and $\bar \phi$ in Eqs. (\ref{VVB}) are indeterminate on the propagation axis, $r=0$. This suggests that the PS of VV beams (\ref{VVB}) comes from the singularity of polarization bases (\ref{PB-para}). To demonstrate this in a different way, we turn to a Stratton vector that is perpendicular to the propagation axis, say $\mathbf{I}^\perp =-\bar{x}$, where $\bar x$ is the unit vector along the $x$ axis.
The wavevector-associated polarization bases it determines take the form of
\begin{subequations}\label{PB-perp}
\begin{align}
  \mathbf{u}^{\perp} =& \frac{\bar{x}-\bar{\rho} \sin^2 \vartheta \cos \varphi
  	-\bar{z} \cos \vartheta \sin \vartheta \cos \varphi}{(1-\sin^2 \vartheta \cos^2 \varphi)^{1/2}}, \\
  \mathbf{v}^{\perp} =& \frac{\bar{y} \cos \vartheta -\bar{z} \sin \vartheta \sin \varphi}
                             {(1-\sin^2 \vartheta \cos^2 \varphi)^{1/2}} ,
\end{align}
\end{subequations}
where $\bar y$ is the unit vector along the $y$ axis.
Of course, they are singular when $\sin \vartheta =1$ and $\cos \varphi =\pm 1$. But they are regular on the propagation axis, $\sin \vartheta =0$.
The polarization vectors described by $\alpha_{1+}$ and $\alpha_{1-}$ in this case become
\begin{equation}\label{PV-perp}
    \mathbf{a}_{1+}^{\perp} =\varpi^{\perp} \alpha_{1+} =\mathbf{u}^{\perp}, \quad
    \mathbf{a}_{1-}^{\perp} =\varpi^{\perp} \alpha_{1-} =\mathbf{v}^{\perp},
\end{equation}
respectively, where
$\varpi^{\perp} =(\begin{array}{cc}
                    \mathbf{u}^{\perp} & \mathbf{v}^{\perp}
                  \end{array}
                 )$.
It is seen from Eqs. (\ref{PB-perp}) that whether the polarization vector $\mathbf{a}_{1+}^{\perp}$ or the polarization vector $\mathbf{a}_{1-}^{\perp}$ has a first-order axial component with respect to $\sin \vartheta$.
Once the strength factor $e(\mathbf{k})$ is sharply peaked at $\vartheta =0$, paraxial approximation applies in accordance with Eq. (\ref{FT}). In the zeroth-order approximation in which $\sin \vartheta \approx 0$ and $\cos \vartheta \approx 1$,
we have
\begin{equation}\label{PV-UPL}
    \mathbf{a}_{1+}^{\perp} \approx \bar x, \quad
    \mathbf{a}_{1-}^{\perp} \approx \bar y.
\end{equation}
So if the factor $f(\vartheta)$ in Eq. (\ref{OAMB}) is sharply peaked at $\vartheta =0$, inserting Eqs. (\ref{OAMB}) and (\ref{PV-UPL}) into Eq. (\ref{FT}) and making use of identity (\ref{iden}), we obtain
\begin{subequations}\label{SB-LP}
\begin{align}
    \mathbf{E}_{1+,m}^{\perp} & \approx i^{m-1} \exp(im \phi) A'_m (r,z) \bar{x}, \\
    \mathbf{E}_{1-,m}^{\perp} & \approx i^{m-1} \exp(im \phi) A'_m (r,z) \bar{y},
\end{align}
\end{subequations}
where a time-dependent factor $\exp(-i \omega_0 t)$ is assumed as before. They are uniformly polarized in $x$ and $y$ directions, respectively, and therefore do not have polarization singularities on the propagation axis.

It is seen from the above discussions that VV beams (\ref{VVB}) and their scalar counterparts (\ref{SB-LP}) have the same Jones vector as well as the same strength factor. They are only different in the Stratton vector. We are thus convinced that the PS of VV beams (\ref{VVB}) on the propagation axis truly comes from the singularity of polarization bases (\ref{PB-para}).

\section{Discussions}

Let us now show that neither of VV beams (\ref{VVB}) is a superposition of two polarization-orthogonal scalar beams. Considering the principle of duality, we only discuss one of them, say TE mode (\ref{VVB2}).
With the help of the relations,
\begin{equation*}
    \bar{r}=    \bar{x} \cos \phi +\bar{y} \sin \phi, \quad
    \bar{\phi}=-\bar{x} \sin \phi +\bar{y} \cos \phi,
\end{equation*}
one can readily convert Eq. (\ref{VVB2}) into
\begin{equation}\label{TE}
    \mathbf{E}^\parallel_{1-,m}
   =\frac{i^{m+1}}{2} e^{i m \phi}
    [A'_{m+1} e^{i \phi}(\bar{x}-i \bar{y}) +A'_{m-1} e^{-i \phi}(\bar{x}+i \bar{y})].
\end{equation}
It is seemingly a superposition of two circularly-polarized scalar beams. But this is not the case at all.
As one particular case of the electric field given by Eqs. (\ref{FT})-(\ref{PB}), expression (\ref{TE}) is an exact solution of free-space Maxwell's equations. It represents the electric field of a true light beam.
Nevertheless, it is well known \cite{Patt-A} that except for plane waves, a light beam cannot be uniformly polarized. In other words, no true uniformly-polarized scalar beams exist.
Therefore, neither constituent of expression (\ref{TE}) can be interpreted as the electric field of a true light beam.
Let us explain this in more detail below.

First of all, it is observed that if the polarization vector $\mathbf a$ were independent of the wavevector $\mathbf k$, electric field (\ref{FT}) would be uniformly polarized.
But according to Eqs. (\ref{QUT}) and (\ref{PB}), the polarization vector must be perpendicular to the wavevector,
\begin{equation}\label{TC}
    \mathbf{k} \cdot \mathbf{a}=0.
\end{equation}
This is in fact the requirement of Maxwell's equation (\ref{ME}).
Secondly, it has been shown by Eqs. (\ref{PV-UPL}) that when the Stratton vector is perpendicular to the propagation axis, the polarization vector in the zeroth-order paraxial approximation can be independent of the wavevector.
We consider, under such a Stratton vector, the polarization vector that is described by
$
\alpha_{\sigma_3}= \frac{1}{\sqrt 2} \bigg(\begin{array}{c}
                                             1         \\
                                             i \sigma_3
                                           \end{array}
                                     \bigg)
$,
the eigenfunction of the Pauli matrix
$
\hat{\sigma}_3 =\bigg(\begin{array}{cc}
                        0 & -i \\
                        i &  0
                      \end{array}
                \bigg)
$
with the eigenvalue $\sigma_3 =\pm 1$, which is given by
\begin{equation*}
    \mathbf{a}^\perp_{\sigma_3}
   =\varpi^\perp \alpha_{\sigma_3}
   =\frac{1}{\sqrt 2}(\mathbf{u}^\perp +i \sigma_3 \mathbf{v}^\perp ).
\end{equation*}
In the zeroth-order paraxial approximation, it tends to be independent of the wavevector,
\begin{equation}\label{PV-UPC}
    \mathbf{a}^\perp_{\sigma_3}
    \approx \frac{1}{\sqrt 2} (\bar{x} +i \sigma_3 \bar{y}).
\end{equation}
If the factor $f(\vartheta)$ in Eq. (\ref{OAMB}) is sharply peaked at $\vartheta =0$, we insert Eqs. (\ref{OAMB}) and (\ref{PV-UPC}) into Eq. (\ref{FT}) and take identity (\ref{iden}) into account to get
\begin{equation*}
    \mathbf{E}^\perp_{\sigma_3,m}
   \approx \frac{1}{\sqrt 2} i^{m-1} \exp(im \phi) A'_m (r,z) (\bar{x} +i \sigma_3 \bar{y}),
\end{equation*}
where a time-dependent factor $\exp(-i \omega_0 t)$ is assumed and $A'_m (r,z)$ is given by Eq. (\ref{A'm}).
It is uniformly circularly-polarized, resembling the constituents of expression (\ref{TE}). However, it no longer satisfies Maxwell's equation (\ref{ME}) because paraxially approximated polarization vector (\ref{PV-UPC}) does not satisfy Eq. (\ref{TC}). It cannot represent the electric field of a true light beam.
This means that neither constituent of expression (\ref{TE}) can be viewed as the electric field of a true light beam.
In a word, the PS of VV beams (\ref{VVB}) cannot be interpreted as the result of a superposition of two polarization-orthogonal scalar beams.

Even so, one may still wonder why TE mode (\ref{VVB2}) in general has a radially polarized component if its PS comes from the singularity of the polarization vector (\ref{PV-TE}) that is purely azimuthally polarized. To answer this question, it is worth mention that the polarization state of light beam (\ref{FT}), when viewed in momentum space, is completely described by the polarization vector $\mathbf a$.
Considering that the polarization vector itself as well as constraint (\ref{TC}) on it does not depend on the time, we introduce a position-space vector function defined by
\begin{equation}\label{PF}
    \mathbf{A} (\mathbf{x})
   =\frac{1}{(2 \pi)^3} \int \mathbf{a}(\mathbf{k}) \exp(i \mathbf{k} \cdot \mathbf{x}) d^3 k .
\end{equation}
As the Fourier integral of the polarization vector, it takes the role of describing the polarization state of light beam (\ref{FT}) in position space. In particular, owing to constraint (\ref{TC}) on the polarization vector, it satisfies $\nabla \cdot \mathbf{A} =0$, the same as Maxwell's equation (\ref{ME}). We will refer to it as the polarization function of light beam (\ref{FT}).
It is remarked that when the Jones vector $\alpha$ is independent of the wavevector $\mathbf k$, polarization function (\ref{PF}) takes the form
$\mathbf{A} (\mathbf{x})= \Pi(\mathbf{x}) \alpha$,
where
\begin{equation*}
    \Pi(\mathbf{x})
   =\frac{1}{(2 \pi)^3} \int \varpi \exp(i \mathbf{k} \cdot \mathbf{x}) d^3 k
   \equiv (\begin{array}{cc}
             \mathbf{U} & \mathbf{V}
           \end{array}
          ).
\end{equation*}
In this case, the two vector functions $\mathbf{U}(\mathbf{x})$ and $\mathbf{V}(\mathbf{x})$ act as the polarization bases in position space. They are orthogonal to each other in the sense
$\int \mathbf{U}^\ast \cdot \mathbf{V} d^3 x =0$.
On this basis, if we further introduce a scalar wavefunction defined by
\begin{equation}\label{SF}
    E(\mathbf{x},t)=\frac{1}{(2 \pi)^{3/2}}
                    \int e(\mathbf{k}) \exp[i(\mathbf{k} \cdot \mathbf{x}-\omega t)] d^3 k,
\end{equation}
we will get from Eq. (\ref{FT})
\begin{equation}\label{CI}
    \mathbf{E}(\mathbf{x},t) =\mathbf{A}(\mathbf{x}) \ast E(\mathbf{x},t),
\end{equation}
where $\ast$ means convolution,
\begin{equation*}
    \mathbf{A}(\mathbf{x}) \ast E(\mathbf{x},t)
   =\int \mathbf{A}(\mathbf{x}-\mathbf{x}') E(\mathbf{x}',t) d^3 x' .
\end{equation*}
It is noted that scalar wavefunction (\ref{SF}) satisfies the wave equation
\begin{equation*}
    \nabla^2 E -\frac{1}{c^2} \frac{\partial^2 E}{\partial t^2} =0,
\end{equation*}
the same as wave equation (\ref{WE}). Moreover, it has the property
\begin{equation*}
    \int |E|^2 d^3 x =\int |\mathbf{E}|^2 d^3 x.
\end{equation*}

Equation (\ref{CI}) states that the electric field of a vector beam can be expressed as the convolution of its polarization function $\mathbf A$ with its scalar wavefunction $E$. Let us look at the polarization function of TE mode (\ref{VVB2}), which is
\begin{equation*}
    \mathbf{A}^\parallel_{1-}
   =\frac{1}{(2 \pi)^3} \int \bar{\varphi} \exp(i \mathbf{k} \cdot \mathbf{x}) d^3 k,
\end{equation*}
in accordance with Eqs. (\ref{PF}) and (\ref{PV-TE}). Taking the rotational invariance of $\bar{\varphi}$ into consideration, we perform the integral in cylindrical coordinates to give
\begin{equation}\label{PF-TE}
    \mathbf{A}^\parallel_{1-} =i \frac{\delta(z)}{2 \pi r^2} \bar{\phi}.
\end{equation}
It is only polarized in azimuthal direction, in perfect agreement with polarization vector (\ref{PV-TE}) in momentum space. We are thus clear that the radially polarized component in TE mode (\ref{VVB2}) results from the convolution of polarization function (\ref{PF-TE}) with the scalar wavefunction, which is given by
\begin{equation*}
    E_m =\frac{1}{(2 \pi)^{3/2}}
         \int e_m (\mathbf{k}) \exp(i \mathbf{k} \cdot \mathbf{x}) d^3 k
        =i^{m-1} \exp(i m \phi) A'_m (r,z) ,
\end{equation*}
in accordance with Eqs. (\ref{SF}) and (\ref{OAMB}), where a time-dependent factor $\exp(-i \omega_0 t)$ is assumed.
Similar discussions can also be made about the consistency of the polarization structure of TM mode (\ref{VVB1}) with polarization vector (\ref{PV-TM}).

\section{Conclusions}

In conclusion, we showed that the PS of VV beams (\ref{VVB}) on the propagation axis comes from the singularity of the polarization bases (\ref{PB-para}). It cannot be interpreted as the result of a superposition of two polarization-orthogonal scalar beams.
We also analyzed, with the help of position-space polarization function (\ref{PF}), the consistency of the polarization structure of VV beams (\ref{VVB}) with polarization bases (\ref{PB-para}).

\begin{acknowledgments}
This work was supported in part by the program of Shanghai Municipal Science and Technology Commission under Grant 18ZR1415500.
\end{acknowledgments}



\end{document}